\documentclass[fleqn,usenatbib]{mnras}

\usepackage{newtxtext,newtxmath}

\usepackage[T1]{fontenc}

\DeclareRobustCommand{\VAN}[3]{#2}
\let\VANthebibliography\thebibliography
\def\thebibliography{\DeclareRobustCommand{\VAN}[3]{##3}\VANthebibliography}

\usepackage{graphicx}	
\usepackage{amsmath}	

\title[Making hot Jupiters in stellar clusters]{Making hot Jupiters in stellar clusters: the importance of binary exchange}

\author[D. Li et al.]{
Daohai Li,$^{1}$\thanks{E-mail: lidaohai@bnu.edu.cn, lidaohai@gmail.com (DL)}
Alexander J. Mustill,$^{2}$
Melvyn B. Davies$^{3}$
and  Yan-Xiang Gong$^{4}$
\\
$^{1}$Department of Astronomy, Beijing Normal University, No.19, Xinjiekouwai St, Haidian District, Beijing, 100875, P.R.China\\
$^{2}$Lund Observatory, Department of Astronomy and Theoretical Physics, Lund University, Box 43, SE-221 00 Lund, Sweden\\
$^{3}$Centre for Mathematical Sciences, Lund University, Box 118, 221 00 Lund, Sweden\\
$^{4}$College of Physics and Electronic Engineering, Taishan University, Taian 271000, China
}

\date{Accepted XXX. Received YYY; in original form ZZZ}

\pubyear{2015}

\begin{document}
\label{firstpage}
\pagerange{\pageref{firstpage}--\pageref{lastpage}}
\maketitle

\begin{abstract}
It has been suggested that the occurrence rate of hot Jupiters (HJs) in open clusters might reach several per cent, significantly higher than that of the field ($\sim$ a per cent). In a stellar cluster, when a planetary system scatters with a stellar binary, it may acquire a companion star which may excite large amplitude von Zeipel-Lidov-Kozai oscillations in the planet's orbital eccentricity, triggering high-eccentricity migration and the formation of an HJ. We quantify the efficiency of this mechanism by modelling the evolution of a gas giant around a solar mass star under the influence of successive scatterings with binary and single stars. We show that the chance that a planet $\in(1,10)$ au becomes an HJ in a Gyr in a cluster of stellar density $n_*=50$ pc$^{-3}$ and binary fraction $f_\mathrm{bin}=0.5$ is about 2\% and an additional 4\% are forced by the companion star into collision with or tidal disruption by the central host. An empirical fit shows that the total percentage of those outcomes asymptotically reaches an upper limit determined solely by $f_\mathrm{bin}$ (e.g., $10\%$ at $f_\mathrm{bin}=0.3$ and 18\% at $f_\mathrm{bin}=1$) on a timescale inversely proportional to $n_*$ ($\sim$ Gyr for $n_*\sim100$ pc$^{-3}$). The ratio of collisions to tidal disruptions is roughly a few, and depends on the tidal model. Therefore, if the giant planet occurrence rate is 10~\%, our mechanism implies an HJ occurrence rate of a few times 0.1~\% in a Gyr and can thus explain a substantial fraction of the observed rate.

\end{abstract}




\section{Introduction}

Most stars form in a cluster together with tens to thousands of siblings \citep{Lada2003}. As a prominent example, our Sun probably originated from a cluster with a few thousand stars \citep{Adams2010}. Such a cluster environment may intuitively seem hostile to planet formation, as the UV radiation from the more massive cluster members may destroy the proto-planetary disc \citep[e.g.,][]{Scally2001,Adams2006,Winter2018,Nicholson2019} and the stellar scattering may disperse either the disc \citep[e.g.,][]{Pfalzner2005,Olczak2012,Vincke2016,PortegiesZwart2016} or the already-formed planets \citep[e.g.][]{Spurzem2009,Malmberg2011,Li2015,Cai2017,Li2019,VanElteren2019}.

However, it turns out for open clusters in general, these effects are mild and the planet formation/survivability within a few tens of au is not likely to be affected by the cluster environment \citep{Laughlin1998,Adams2001,Adams2006,Malmberg2011,Hao2013,Li2015,Cai2017,Fujii2019,Li2019,Li2020a,Li2020c}. An obvious example is again our solar system that originated from a sizeable cluster but managed to retain objects out to at least tens of au. Thus one would expect that planets around stars in open clusters should look similar to those orbiting field stars. The observations are still sparse with only a dozen planets found in clusters \citep{Meibom2013,Quinn2012,Quinn2014,Brucalassi2016,Obermeier2016,Ciardi2017,Rizzuto2018,Livingston2019} but the data do not seem to disagree with this inference \citep[e.g.,][]{Meibom2013,Brucalassi2017,Takarada2020}.

Nonetheless, one exception may be the hot Jupiters (HJs) which seem tentatively more populous in (some) open clusters. A radial velocity survey by \citet{Quinn2012} reported the discovery of two HJs in the metal rich ([Fe/H]$\sim$0.19) 600-Myr-old Praesepe cluster and an occurrence rate of 3.8~\% was derived. The detection of an HJ in the Hyades cluster (also metal rich with [Fe/H]$\sim$0.13 and about 600 Myr old) was made by \cite{Quinn2014} and combining the previous (non-detection) result \citep{Paulson2004} the authors estimated that the average HJ occurrence rate in Praesepe and Hyades was 2.0~\%. However, it is well known that the HJ/giant planet occurrence rate in the field correlates with the host star's metallicity \citep[e.g.,][]{Gonzalez1997}. After correcting for the solar metallicity, the derived occurrence rate of HJs for the two open clusters became 1~\% \citep{Quinn2014}, in good agreement with that of the field \citep[$\sim1.2\%$; e.g.,][]{Wright2012}. Another radial velocity survey by \citet{Brucalassi2014,Brucalassi2016} of the solar-metallicity and solar-age cluster M67 yielded 3 HJs, leading to an occurrence rate of 5.6 and 4.5~\% considering single star hosts and for any star, respectively. More recently, \citet{Takarada2020} looked into close-in giant planets in the young 100-Myr-old, solar-metallicity open cluster Pleiades and their non-detection has given rise an upper limit of 11~\% for the HJ occurrence rate in that cluster.

It therefore seems that the HJ occurrence rate in open clusters is no smaller than that of the field, and in some occasions appreciably higher. This is, however, under debate, and more data are needed.

The formation of HJs has recently been reviewed by \citet{Dawson2018}. There are three competing theories: in-situ, disc migration, and high eccentricity (high-$e$) migration. In the former two cases, the HJs form in the presence of the gaseous disc so the planet's eccentricity probably remains low because of disc damping while in the latter scenario, the planet's eccentricity is highly excited, leading to a very small pericentre distance; then strong tidal interactions are activated, which shrink and circularise the planet's orbit, forming an HJ. Notably, four out of the six HJs discovered in open clusters as discussed above have moderate eccentricities $\gtrsim0.1$, lending support to high-$e$ migration formation. Mechanisms able to excite high eccentricities include the planets' interaction \citep[either direct scattering or secular forcing; e.g.,][]{Rasio1996,Wu2011} or perturbation by a distant companion star/planet \citep[e.g.,][]{Wu2003,Fabrycky2007,Malmberg2007a,Naoz2011} through the von Zeipel-Kozai-Lidov mechanism \citep[ZKL][]{Zeipel1909,Kozai1962,Lidov1962}.

As mentioned above, in an open cluster, the planetary system's configuration is not expected to be modified by the environment, but perhaps the cluster can boost the formation of HJs via nudging the much further-out stellar companion \citep[or planets, see][ and Section \ref{sec-dis} for a discussion]{Wang2020a,Rodet2021a,Wang2022}. But how does a planetary system acquire such a companion star in the first place?

Stellar binary-single scattering might be a solution. When a stellar binary scatters with a planetary system (effectively a single star as far as the stellar dynamics are concerned), the latter may exchange with a component of the binary and a new binary composed of the planetary system and the other original component forms \citep[e.g.][]{Heggie1975,Hut1983}. \citet{Li2020c} showed that when scattering with a binary star, a planetary system may, while remaining intact during the scattering, acquire a distant companion star. For the Sun-Jupiter system, this scenario happens at a rate an order of magnitude higher than that of the planet's ejection. For an open cluster with a stellar number density of 50 pc$^{-3}$ and a binarity (the fraction of binary systems among all systems) of 0.5, the Sun-Jupiter system has a chance of 10\% to obtain a companion within 100 Myr. \citet{Li2020c} also estimated that half of those so-acquired companion stars can activate ZKL cycle in the planet's orbital evolution. But how efficiently can this process initiate high-$e$ migration and create HJs? This is the question we want to answer in this work. 

The paper is organised as follows. In Section \ref{sec-met}, we describe the simulation strategy, detailing the modelling of the scattering and the tidal dissipation model. The relevant timescales are compared in Section \ref{sec-timescale}. We present two sets of test simulations in Section \ref{sec-test} where a Jupiter mass planet is placed at 5 au. In Section \ref{sec-pop} we show our population synthesis simulations where the planets' initial orbits are taken from the observed population and the cluster properties are varied within reasonable ranges. We discuss the implications and present the main results in Section \ref{sec-dis} and Section \ref{sec-con}, respectively.

\section{Method}\label{sec-met}
In open clusters, a planetary system may encounter more than one stellar system \citep[e.g.,][]{Malmberg2007} and when not interacting with scattering stars, the system evolves on its own. In the following, we first describe how the stellar scatterings are generated and then how the simulation is designed.

\subsection{Creation of the scatterings}
We refer to the stellar systems that scatter with the planetary system as scatterer and they can be a single star or a binary. The rate at which the planetary system encounters a scatterer can be estimated with 
\begin{equation}
\label{eq-scat-rate}
\Gamma=n_* \sigma v_\mathrm{inf},
\end{equation}
where $n_*$ is the number density of the stellar systems in the cluster (including both single and binary systems), $v_\mathrm{inf}$ the relative velocity of the scattering and $\sigma$ the encounter cross section. While typical values for the former two can be found in standard references \citep[e.g.,][]{Binney2008}, $\sigma$ will be defined here and therefore prescribes $\Gamma$. We would like to have a large enough $\sigma$ such that no important encounters will be missed but also one that is not too large to avoid the numerous weak unimportant scatterings overwhelming our simulation. Therefore, $\sigma$ has to be chosen by the result of the encounter.

The outcome of a scattering event critically depends on how close the planetary system and the scatterer get and this distance can be linked to the scatterer mass, impact parameter $b$, and $v_\mathrm{inf}$ via gravitational focusing. \citet{Li2020c} performed scattering experiments between the Sun-Jupiter system and a stellar binary/single star, varying the stellar mass, orbital semimajor axis (if the scatterer is a binary), and $v_\mathrm{inf}$. They found that the maximum impact factor $b_\mathrm{max}$ at which an encounter might still lead to disruption/exchange events can be approximated by (the top equation in their figure 2 and after a little algebraic manipulation)
\begin{equation}
\begin{aligned}
\label{eq-b-max}
&\log {b_\mathrm{max}(m_\mathrm{tot},a_\mathrm{tot},v_\mathrm{inf})\over 1 \,\mathrm{au}}=2.04+0.51\log {m_\mathrm{tot}\over 1 \,\mathrm{M}_\odot}\\
&+0.49\log {a_\mathrm{tot}\over 1 \,\mathrm{au}} -1.00 \log {v_\mathrm{inf}\over 1 \,\mathrm{km\,s}^{-1}},
\end{aligned}
\end{equation}
in which $a_\mathrm{tot}=a_\mathrm{pl}+a_\mathrm{bin}$ is the sum of the planetary semimajor axis and that of the scattering binary in au (if the scatter is a single star, $a_\mathrm{bin}=0$); $m_\mathrm{tot}$ is total mass of all the objects in Solar masses (M$_\odot$); and $v_\mathrm{inf}$ is in km~s$^{-1}$. Therefore, in order not to miss any important encounter, the cross section has to be at least $\pi b^2_\mathrm{max}$ which depends on the property of the scatterer and the planetary system as above.

Now we describe how the scatterers are created. Be the scatterer a single star or a binary stellar system, the stellar mass is drawn independently from the initial mass function by \citet{Kroupa2001} with a lower limit of 0.1 M$_\odot$ and an upper limit of 10 M$_\odot$. For a binary scatterer, the relative orbit is created following the observed distribution of solar type binaries in the field \citep{Duquennoy1991,Raghavan2010} as done in \citet{Li2020c}. As tight binaries behave effectively like a single star \citep{Li2020c}, in our simulation, if the binary semimajor axis $a_\mathrm{bin}<1$ au, the two components are merged as a single object. The upper limit for the $a_\mathrm{bin}$ has been set to 1000 au, roughly where the binary becomes soft in an open cluster (see Figure \ref{fig-timescales} below). And, $v_\mathrm{inf}$ is drawn from a Maxwellian distribution with a mean of 1 km~s$^{-1}$. Finally, we draw $b\in (0,b_\mathrm{ext})$ such that the probability distribution function is proportional to $b$. Here the constant $b_\mathrm{ext}$ is the extreme value for $b$ such that any encounter with $b>b_\mathrm{ext}$ cannot be important in our simulations (so $b_\mathrm{ext}\ge b_\mathrm{max}$ for any scattering parameter) and its determination will be discussed next.

The value of $b_\mathrm{ext}$ is effectively the same as the largest possible $b_\mathrm{max}$ required by the combination of the highest scatterer mass, largest $a_\mathrm{bin}$, and smallest $v_\mathrm{inf}$. We consider a binary scatterer with the two components each of $10\mathrm{M}_\odot$, $a_\mathrm{bin}=1000$ au, and $v_\mathrm{inf}=0.01$ km~s$^{-1}$. Though $v_\mathrm{inf}$ can be arbitrarily small, we note 0.01 km~s$^{-1}$ $\approx$ 0.01 pc~Myr$^{-1}$ and it takes such an encounter a few hundreds of Myr to traverse a typical open cluster of a few pc so the chance of these slow encounters is small. With these extreme values, $b_\mathrm{ext}\sim8$ pc and thus $\sigma_\mathrm{ext}\sim2\times10^2$ pc$^2$. The scattering frequency is consequently $\Gamma=200n_*$ Myr$^{-1}$ (where $n_*$ is in pc$^{-3}$ and we have taken $v_\mathrm{inf}=1$ km~s$^{-1}$), suggesting that the scatterings are extremely frequent. But a significant proportion of the encounters will not do anything appreciable either to the planetary system or to the scatterer (if binary) and do not need to be considered because their $b$ is larger than the respective $b_\mathrm{max}$. Therefore, upon creating a scatterer with its time of encounter, mass, orbit, $b$ and $v_\mathrm{inf}$, if $b>b_\mathrm{max}$, we simply omit this encounter and proceed to the next.

If the planetary system has an outer stellar companion, we will need to account for the evolution of the companion orbit during the encounters as well. Hence, in Equation \eqref{eq-b-max}, now $a_\mathrm{tot}=a_\mathrm{com}+a_\mathrm{bin}$, the sum of the binary and the companion semimajor axes.

Those scattering events are simulated in a way similar to {\small FEWBODY} \citep{Fregeau2004}. On initiation, with the scatterer mass, $v_\mathrm{inf}$ and $b$, we analytically move it to a distance such that the tidal perturbation on the star-planet/stellar binary system relative to their internal forcing is smaller than $10^{-4}$. During the scattering, we look for stable binary/triple systems recursively and the scattering is deemed finished if all triples are stable and/or the tidal perturbation on any binary by any other object is small again.

\subsection{Simulation strategy}\label{sec-met}

In between the stellar encounters, the planetary system (if there is no companion star) evolves on its own and this can be tracked analytically following a two-body prescription. If there is a companion star or if the planet's pericentre distance $r_\mathrm{peri,pl}$ is small so general relativity (GR) and tides are important, the planet's orbit is propagated numerically (GR and tides are also included during the scattering if needed). In our implementation, the GR effect is approximated by the leading order post-Newtonian potential \citep{Kidder1995} following \citet{Bolmont2015}. The equilibrium tidal model \citep{Hut1981} is adopted in this work as done in \citet{Bolmont2015}. In the formation of HJs, however, the planetary orbit can be extremely eccentric and thus dynamical tides involving different modes in the planet's/star's oscillation become important. Here we follow \citet{Beauge2012} and simply vary the tidal quality factor $Q_\mathrm{tide}$ according to $r_\mathrm{peri,pl}$ and the planet's eccentricity $e_\mathrm{pl}$ to mimic the said effect
\begin{equation}
\label{eq-tid-q}
Q_\mathrm{tide}=10^{200e_\mathrm{pl}^2({r_\mathrm{peri,pl}\over 1 \mathrm{au}}-0.022)}Q_\mathrm{tide,0},
\end{equation}
where $Q_\mathrm{tide,0}$ is $10^7$ and $5\times10^6$ for the host star and the planet respectively. \citet{Beauge2012} found the above formula able to reproduce the migration and circularisation timescale predicted by the dynamical tidal model of \citet{Ivanov2011} fairly well. Equation \eqref{eq-tid-q} allows for an easy correction for the qualitative features of dynamical tides within the framework of the equilibrium tidal model. When $r_\mathrm{peri,pl} < 0.022$ au, $Q_\mathrm{tide}<Q_\mathrm{tide,0}$ and in the meantime, if $1-e_\mathrm{pl}\ll 1$, $Q_\mathrm{tide}$ can be much smaller than $Q_\mathrm{tide,0}$ and tidal dissipation is efficient. Otherwise, $Q_\mathrm{tide}>Q_\mathrm{tide,0}$ and tides are ineffective. This set of tidal parameter is adopted throughout the work unless explicitly stated otherwise. In a subset of the simulations, we have also introduced an enhanced tidal model, where $Q_\mathrm{tide,0}$ is reduced by a factor of ten compared to the values above to mock more efficient tidal damping models \citep[e.g.,][]{Wu2018}.

Tidal evolution has to do with the exchange of energy and angular momenta of the orbital motion and spins of the star/planet. While the tidal deformation in the planet is the main driver for the orbital circularisation, that in the star is able to further modify the orbit afterwards. As the planet's spin angular momentum is much less than that of its orbital motion, the orbital angular momentum is effectively conserved in the first stage.

In this work, we stop the propagation of a planetary system once the planet's apocentre distance $r_\mathrm{apo,pl}$ drops below 1 au in order to save computational time. We deem that beyond this point, the formation of an HJ is unlikely to be disrupted by further encounters or the small planet orbit makes it immune from further external perturbation and the formation of an HJ is impossible. Therefore, until this point, the planet's orbit is still highly eccentric and only the planetary tide is important in our simulation, enabling a simple handling of the spins of the two objects. The planet, due to its small mass and physical radius, carries a much smaller spin angular momentum than that of the orbital motion and its spin is aligned and (pseudo-)synchronised with the orbital angular velocity at pericentre on a timescale much shorter than that of orbital evolution \citep[e.g.,][]{Hut1981,Correia2009}. Hence, we simply let the spin of the planet be in that status in our simulation \citep[e.g.,][]{Hamers2017a}. Though during the ZKL cycles, the stellar spin may evolve in a very complicated manner \citep{Storch2014}, it is not expected to affect the orbital dynamics during the planet's orbital shrinkage but may play a significant role latter on \citep[e.g.,][]{Fabrycky2007}, beyond the scope of this work. We let the stellar spin be the current solar value and align it with the initial orbital plane of the planet.

Both GR and tides are only effective when $r_\mathrm{peri,pl}$ is small. In our code, the two are activated only if $r_\mathrm{peri,pl}<0.05$ au (ten solar radii).

The code checks the planet's orbital elements at the beginning and at the end of all scatterings and also routinely does so not during a scattering. If the planet's apocentre distance drops below 1 au and at the same time, the pericentre distance is below 0.02 au, we deem that an HJ forms.

Finally, the Bulirsch-Stoer integrator available in the {\small MERCURY} package \citep{Chambers1999} is adopted for propagating the state vectors of the objects using an error tolerance of $10^{-12}$. And collisions between the objects are also detected using the subroutines in {\small MERCURY}. The planetary system is followed for 1 Gyr and the simulation is stopped if the planet becomes an HJ or does not orbit the original host anymore.
\section{Timescales}\label{sec-timescale}
Numerous authors have examined high-$e$ migration in different context \citep[e.g.,][and see \citet{Naoz2016} for a review]{Wu2003,Fabrycky2007}. We would like to briefly review some of the relevant  timescales.

Under the point mass Newtionian gravity assumption, a companion star may excite the planet's orbital eccentricity to arbitrarily large eccentricities via the ZKL mechanism \citep[e.g.,][]{Ford2000,Takeda2008}. This picture changes as the ZKL cycles may be suppressed by other perturbations, for example, the short-range forces GR and/or tides as discussed in this work, exerting faster precession in the planet's orbit \citep[e.g.,][]{Naoz2013a,Liu2015,Naoz2016}. The respective expressions for these timescales are as follows. That of the ZKL cycle is \citep{Antognini2015}
\begin{equation}
\mathrm{ZKL}\sim{8\over 15\pi}{m_\mathrm{host}+m_\mathrm{com}\over m_\mathrm{com}}{P^2_\mathrm{com} \over P_\mathrm{pl}}(1-e^2_\mathrm{com})^{3/2},
\end{equation}
where $m_\mathrm{host}=1$ M$_\odot$ is the mass of the planetary host star and the companion mass $m_\mathrm{com}=0.3$ M$_\odot$; $P_\mathrm{com}$ and $P_\mathrm{pl}$ are the orbital periods of the companion star and the planet, respectively. The timescale of the leading order GR precession is \citep{Naoz2013a}
\begin{equation}
\mathrm{GR}\sim{2\pi\over 3}{a^{5/2}_\mathrm{pl} c^2 \over (G m_\mathrm{host})^{3/2}}(1-e^2_\mathrm{pl}),
\end{equation}
where $c$ is the speed of light and $G$ the gravitational constant. Finally, the tidal bulges raised on the planet lead to orbital precession on a timescale \citep{Naoz2016}
\begin{equation}
\mathrm{Tide}\sim{m_\mathrm{pl}a^{13/2}_\mathrm{pl} \over G^{1/4} k_2 m_\mathrm{host}(m_\mathrm{host}+m_\mathrm{pl}) R^5_\mathrm{pl}}{(1-e^2_\mathrm{pl})^5\over1+{3\over2}e^2_\mathrm{pl}+{1\over8}e^4_\mathrm{pl}},
\end{equation}
where $k_2=0.38$ is the planet's Love number \citep[the same as the Jovian value][]{Bolmont2015} and $R_\mathrm{pl}=7\times10^4$ km its physical radius.

\begin{figure}
\includegraphics[width=\columnwidth]{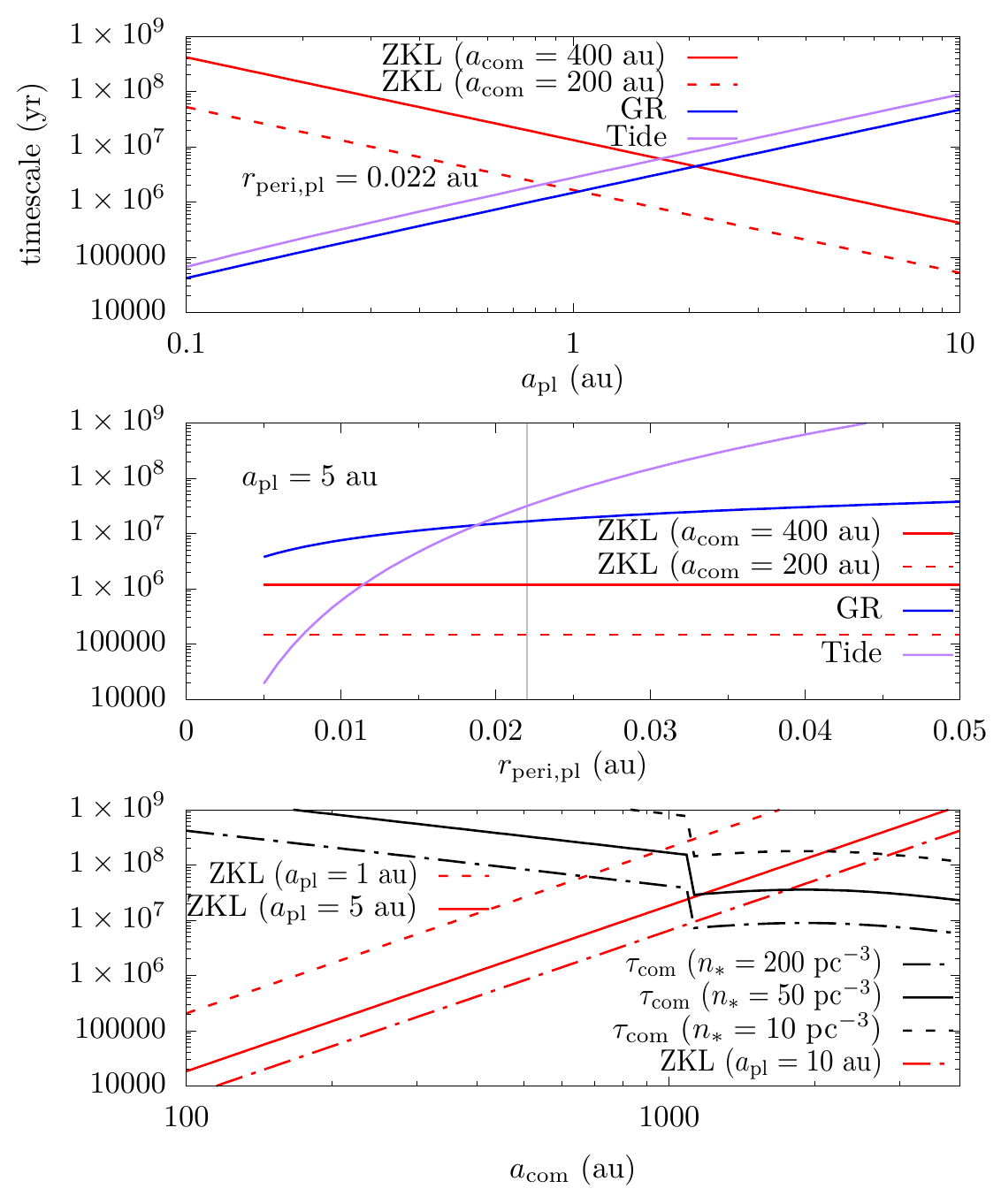}
\caption{Timescales of orbital precession caused by different mechanisms. The three panels show the timescales for the planet's orbital precession owing to the ZKL mechanism of a companion star in red (different line type for different $a_\mathrm{pl}$ and $a_\mathrm{com}$), the GR effect in red, tides in purple, and the lifetime of the companion star in black (different line type for different $n_*$).}
\label{fig-timescales}
\end{figure}

In the top panel of Figure \ref{fig-timescales}, we show the precession timescales by the leading order ZKL effect in red, GR in blue and tides in purple \citep[the relevant expressions are taken from][]{Antognini2015,Naoz2013a,Naoz2016} as a function of $a_\mathrm{pl}$ fixing the planet pericentre $r_\mathrm{peri,pl}=0.022$ au (where tidal effects becomes efficient in our model; see Equation \eqref{eq-tid-q}). The companion orbit has been fixed at $a_\mathrm{com}=400$ (solid line) or 200 au (dashed) and $e_\mathrm{com}=0.7$ (see Figure \ref{fig-hj_com} below for companion orbits from the simulations). Reading from the plot, the ZKL timescale is inversely dependent on $a_\mathrm{pl}$ while those of GR and tide positively. For $a_\mathrm{pl}\gtrsim 1-2$ au, the planet's orbital precession is mainly driven by the companion star. Otherwise, those by GR and tides take over. Whereas the ZKL timescale is insensitive to the $e_\mathrm{pl}$, both GR and tides depend critically on it (or $r_\mathrm{peri,pl}$) and the larger the $r_\mathrm{peri,pl}$, the longer the latter two timescales. This means for $r_\mathrm{peri,pl}>0.022$ au, the ZKL effect prevails (at least for $a_\mathrm{pl}\gtrsim 1-2$ au) and can excite $e_\mathrm{pl}$ to the point where tides are important.

In middle panel, these timescales are shown as a function of $r_\mathrm{peri,pl}$ (or equivalently $e_\mathrm{pl}$), now fixing $a_\mathrm{pl}$ at 5 au. The timescale of the ZKL mechanism does not depend on $e_\mathrm{pl}$/$r_\mathrm{peri,pl}$ and is shown as the red horizontal lines. Both GR and tides depend on $r_\mathrm{peri,pl}$ and the latter more steeply. For $r_\mathrm{peri,pl}\gtrsim 0.012$ au, ZKL timescale is the shortest among the three. This strengthens the above argument and means that for $a_\mathrm{pl}=5$ au, $r_\mathrm{peri,pl}$ may be lowered to $\sim$ 0.012 au by the ZKL mechanism uninterruptedly.

Embedded in an open cluster, the companion star, once obtained by the planetary system, is subject to further stellar scattering and may be thus stripped. The lifetime of the central host--companion binary can be estimated through
\begin{equation}
\tau_\mathrm{com}\sim{1\over n_* \sigma_\mathrm{disp} v_\mathrm{inf}},
\end{equation}
where the scattering velocity $v_\mathrm{inf}=1$ km~s$^{-1}$ and cluster's stellar density $n_*\in(10,200)$ pc$^{-1}$. And $\sigma_\mathrm{disp}$ is the cross section for the companion star's disruption from the planetary host star. When the binary is hard, the term disruption means exchange so the original host-companion pair ceases to exist and the relevant expression can be found in \citet{Bacon1996}; when then binary is soft, disruption additionally includes ionisation and we refer to \citet{Hut1983} for the formulae. From those expressions, $\sigma_\mathrm{disp}$ depends on $m_\mathrm{host}=1$ M$_\odot$, $m_\mathrm{com}=0.3$ M$_\odot$, the scatterer mass $m_\mathrm{scat}=0.3$ M$_\odot$, $a_\mathrm{com}$, $e_\mathrm{com}=0.7$ and $m_\mathrm{vinf}=1$ km~s$^{-1}$. The bottom panel of Figure \ref{fig-timescales} displays this timescale in black compared to that of the ZKL mechanism in red as a function of $a_\mathrm{com}$ for different $a_\mathrm{pl}$ and $n_*$. The former behaves discontinuously at $a_\mathrm{com}\sim 1000$ au where the hard--soft boundary lies. Importantly, the figure shows clearly that for the $a_\mathrm{pl}\in(1,10)$ au and $n_*\in(10,200)$ pc$^{-3}$, the companion star a few hundreds of au apart can enforce full ZKL cycles before it is removed by stellar scattering.

We have shown in our model, a companion star at a few hundreds of au can, via the ZKL mechanism, pump the planet's eccentricity high enough to trigger efficient tidal evolution, not interrupted by tides/GR or scattering stripping of the companion. But we caution that the planet's tidal orbital shrinkage and dynamical decoupling from the companion star may take many ZKL cycles \citep[e.g.,][and see Figure \ref{fig-aei_105_1} below]{Wu2003,Fabrycky2007,Anderson2016} so the companion star has to survive much longer than the ZKL timescale to  this process. Therefore, the requirement that the ZKL timescale is shorter than the companion lifetime is only necessary but not sufficient for high-$e$ migration. In the following, we present a few concrete examples. 

\section{Test simulations}\label{sec-test}
To validate our code, we first perform two sets of test simulations. In these simulations, the planet host is the Sun. In the first, the planet is much like our Jupiter with a circular orbit at 5 au from the central host. The normal tidal model (with $Q_\mathrm{tide,0}=10^7$ and $5\times10^6$ for the host star and the planet) is adopted. A total of 3000 runs are done. This set of simulations is referred to as our ``Jupiter'' run. In the second set, the only difference is that the tidal quality factors are reduced by a factor of ten (with $Q_\mathrm{tide,0}=10^6$ and $5\times10^5$) and called ``JupEnT'' (Jupiter enhanced tides). In both, the cluster property is $n_*=50$ pc$^{-3}$ and binarity $f_\mathrm{bin}=0.5$ (number of binary system divided by the sum of single and binary systems). The simulation parameters are listed in Table \ref{tab-set}.

\begin{table*}
\centering
\caption{Initial setup of the simulations. The first column is the simulation designation, the second the cluster stellar number density $n_*$, the third the binarity $f_\mathrm{bin}$ (the total number of binary star systems divided by the sum of the number of binary and single star systems), the fourth and the fifth the planet's orbital semimajor axis $a_\mathrm{pl}$ and eccentricity $e_\mathrm{pl}$, the sixth the tidal model, and the last the number of runs.}
\label{tab-set}
\begin{tabular}{ccccccc} 
\hline
sim ID &$n_*$ (pc$^{-3}$)& $f_\mathrm{bin}$ & $a_\mathrm{pl}$ (au) & $e_\mathrm{pl}$ & tidal model & $\#_\mathrm{run}$ \\
\hline
Jupiter & 50 & 0.5 & 5 & 0&normal & 3000\\
JupEnT & 50 & 0.5 & 5 & 0 &enhanced& 3000\\
\hline
Nominal & 50 & 0.5 & 1-10 & 0-0.95 &normal& 30000\\
LowDen & 10 & 0.5 & 1-10 & 0-0.95 &normal& 3000\\
HighDen & 200 & 0.5 & 1-10 & 0-0.95 &normal& 3000\\
LowBin & 50 & 0.1 & 1-10 & 0-0.95 &normal& 3000\\
HighBin & 50 & 0.9 & 1-10 & 0-0.95 &normal& 3000\\
\hline
\end{tabular}
\end{table*}

\subsection{Example planet evolution}

\begin{figure}
\includegraphics[width=\columnwidth]{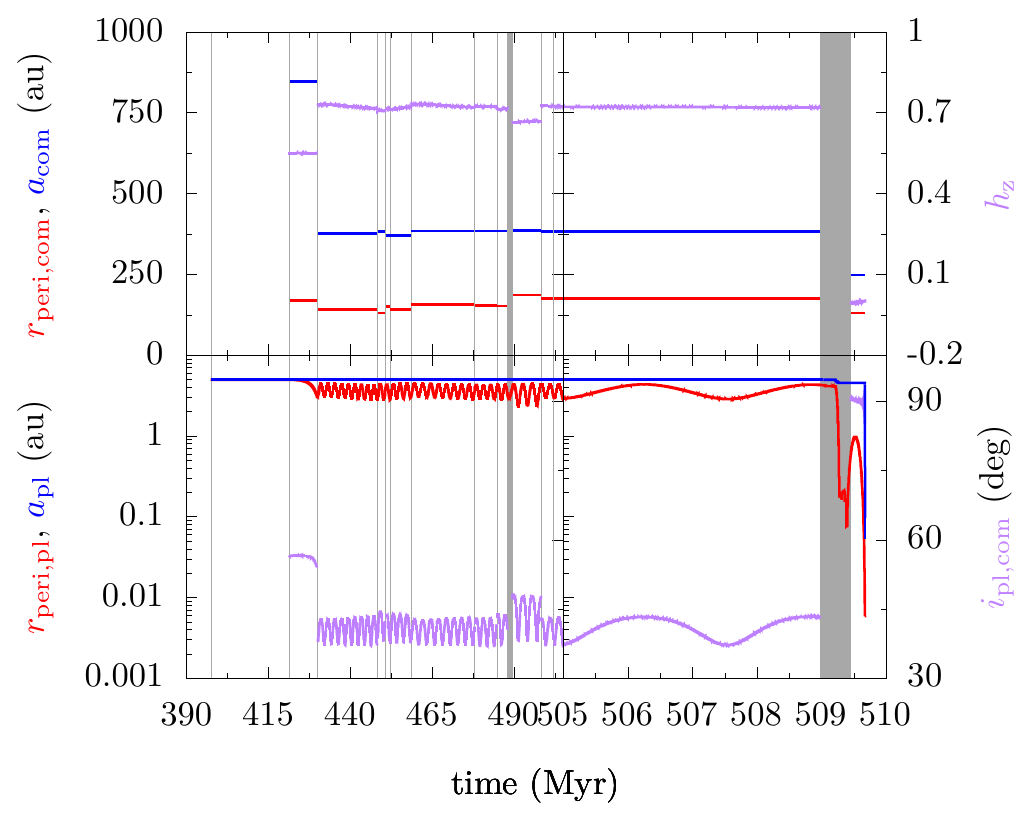}
\caption{Formation of an HJ via HJ\_ZKL. The bottom panel shows the time evolution of the planet's $r_\mathrm{peri,pl}$ (red) and $a_\mathrm{pl}$ (blue) in the left $y$-axis and $i_\mathrm{pl,com}$ (purple) in the right $y$; the top panel shows the companion's $r_\mathrm{peri,com}$ (red) and $a_\mathrm{com}$ (blue) in left $y$-axis and the planet's $h_\mathrm{z}$ (purple) in right $y$. The shaded regions represent ongoing stellar scattering; quantities related the companion's orbit are not shown as it can be not well defined if the scattering is strong.}
\label{fig-aei_105_1}
\end{figure}

Figure \ref{fig-aei_105_1} shows the formation of an example HJ from the Jupiter run. In the plot, the grey regions represent ongoing stellar scattering. Before 400 Myr, the system experiences only one scattering as without a companion, the system is only 5 au wide so a scatterer has to come with a very small impact parameter according to Equation \eqref{eq-b-max} to be potentially important but these are rare. This scattering event does not lead to appreciable changes in the planet's $r_\mathrm{peri,pl}$ (red, bottom panel, left ordinate) or $a_\mathrm{pl}$ (blue, bottom panel, left ordinate). Another scattering with a stellar binary occurs at 420 Myr where the planetary system acquires a companion with pericentre distance $r_\mathrm{peri,com}=170$ au (red, top panel, left ordinate) and $a_\mathrm{com}=850$ au (blue, top panel, left ordinate); and the planet's inclination with respect to the companion's orbital plane is $i_\mathrm{pl,com}=57^\circ$ (purple, bottom panel, right ordinate). Now ZKL cycles are activated in the planet's orbit shown as the phase-correlated oscillations in $r_\mathrm{peri,pl}$ and $i_\mathrm{pl,com}$. The purple line in the top panel shows the planet's normalised vertical orbital angular momentum $h_\mathrm{z}=\sqrt{1-e^2_\mathrm{pl}}\cos i_\mathrm{pl,com}$ relative to the companion's orbital plane with the right $y$-axis which is a conserved quantity in the lowest order ZKL theory. As expected, $h_\mathrm{z}$ is quasi-constant, at least before the next stellar scattering.

The wide orbit of the companion means that more distant scatterings need to be taken into account, indicated by the increase in the number of the grey regions after the acquisition of the companion. During each of these (distant) scatterings, the planet's orbit is not affected but the companion's $r_\mathrm{peri,com}$ and $a_\mathrm{com}$ as well as its inclination $i_\mathrm{pl,com}$ can be instantly altered. This also changes $h_\mathrm{z}$ (because the reference plane changes) so after each scattering, the planet's orbital elements evolve with a new pattern so $r_\mathrm{peri,pl}$ and $i_\mathrm{pl,com}$ reach different extrema.

Finally, during the scattering at 509 Myr, $a_\mathrm{com}$ becomes 380 au, and $i_\mathrm{pl,com}$ reaches almost $90^\circ$. Immediately after this encounter, $r_\mathrm{peri,pl}$ is driven to $<0.01$ au. Then tidal effects quickly shrink the orbit to completely within 1 au during the first minimum of $r_\mathrm{peri,pl}$ so an HJ has formed and the simulation is stopped. This type of outcome is called HJ\_ZKL (formation of HJ by a companion star). It is common (60~\% of all the HJ\_ZKL cases) for a companion star to experience stellar scatterings before it leads to the formation of an HJ.

Is the high-$e$ migration process enhanced by these scatterings? To answer this question we perform a simple test. For each system of the outcome HJ\_ZKL, we have taken a snapshot of it the moment the system acquires the companion star that later leads to the HJ formation. From this snapshot, the system is propagated with the companion star in isolation without any scattering until 1 Gyr. It turns that an HJ only forms in 40~\% of these simulations. This suggests that the scatterings between the companion star and other stars in the cluster have boosted the HJ formation.

\begin{figure}
\includegraphics[width=\columnwidth]{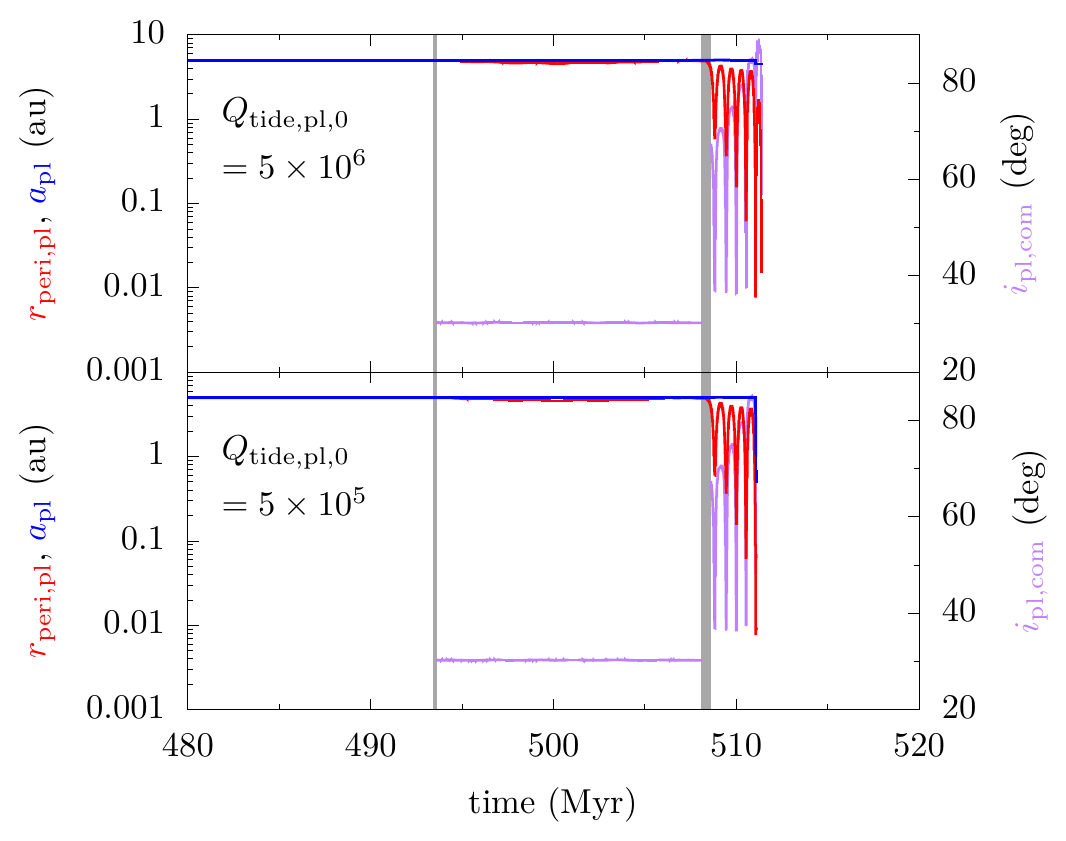}
\caption{Time evolution of a planet that end up in the fates COL\_ZKL and HJ\_ZKL with different tidal $Q$. The left ordinate marks the planet's $r_\mathrm{peri,pl}$ (red) and $a_\mathrm{pl}$ (blue) and the right ordinate $i_\mathrm{pl,com}$ (purple). The top panel shows the case of COL\_ZKL where the planet's $Q_\mathrm{tide,pl,0}=5\times10^6$ and the bottom panel of HJ\_ZKL where $Q_\mathrm{tide,pl,0}=5\times10^5$; all the other parameters are the same in the two.}
\label{fig-aei_1_2}
\end{figure}

Not all planets where ZKL cycles of appreciable amplitudes are enabled turn into HJs. For the vast majority of such planet orbits, the maximum $e_\mathrm{pl}$ is simply not large enough such that $r_\mathrm{peri,pl}>0.022$~au and efficient tidal dissipation is never activated. And for another substantial fraction, the planet is driven into the central host by the companion -- tides- and GR-induced orbital precession is outpaced by that of the ZKL effect so the latter goes untamed. The top panel of Figure \ref{fig-aei_1_2} shows such an example. At 494 Myr into the simulation, the planetary system obtains a companion star of $a_\mathrm{com}=700$ au and $e_\mathrm{com}=0.92$ and $i_\mathrm{pl,com}=30^\circ$. Subsequently around 508 Myr, a scattering changes the companion orbit to $a_\mathrm{com}=720$ au, $e_\mathrm{com}=0.96$ and $i_\mathrm{pl,com}=67^\circ$. Now the ZKL cycles are greatly amplified and after several cycles, noticeable higher-order effects manifest by driving down the extreme $r_\mathrm{peri,pl}$ in each successive ZKL cycle \citep[e.g.,][]{Naoz2011}. Then at 501 Myr, when $r_\mathrm{peri,pl}$ reaches 0.01 au, $a_\mathrm{pl}$ drops by $\sim$ 10\% by tides. During the subsequent dip of $r_\mathrm{peri,pl}$, the planet dives into the central host directly before tides are able to do anything. We note that the planet may be tidally disrupted by the star en route to a collision. But the tidal disruption limit for Jupiter around the Sun is about a few solar radii so we do not detect tidal disruption and generally call those collisions. The outcome of the collision with the central star as a result of the ZKL effect by the companion is referred to as COL\_ZKL.

\subsection{Statistics}\label{sec-jup-res}
We count the number of planets that have different fates and show their percentages in Table \ref{tab-res}. The details of our simulations are presented in Section \ref{sec-met}. The second column shows the percentage of the outcome HJ\_ZKL, the third column HJ\_SCAT (an HJ forms where the small $r_\mathrm{peri,pl}$ is established during the scattering, but not forced by a bound companion star), the fourth column COL\_ZKL, and the fifth column EJEC (Jupiter turns into a free floating planet without a host star).

\begin{table*}
\centering
\caption{Percentage of planets with different fates. The first column shows the ID of the simulation set; from the second to the fifth, those for HJ\_ZKL (formation of HJ via the ZKL mechanism by a companion), HJ\_SCAT (formation of HJ where the small pericentre distance is achieved directly during the scattering), COL\_ZKL (collision forced by the ZKL mechanism by a companion), and EJEC (ejection) are shown. The errors are 1-$\sigma$ dispersion from random resampling. The nominal set has 30000 runs while the others have 3000 for each. In Section \ref{sec-dep}, the sum of HJ\_ZKL and COL\_ZKL will be also referred to as ZCT\_ZKL.}
\label{tab-res}
\begin{tabular}{ccccc} 
\hline
sim ID & HJ\_ZKL & HJ\_SCAT & COL\_ZKL&EJEC \\
\hline
Jupiter & $2.43_{-0.26}^{+0.30}$ & $0$ & $5.80_{-0.41}^{+0.47}$&$23.6_{-0.7}^{+0.8}$ \\
JupEnT & $4.27_{-0.40}^{+0.30}$ & $0.0667_{-0.0667}^{+0.0333}$ & $4.17_{-0.37}^{+0.33}$&$23.4_{-0.8}^{+0.7}$ \\
\hline
Nominal & $2.43_{-0.09}^{+0.09}$ & $0.127_{-0.017}^{+0.023}$ &$3.64_{-0.11}^{+0.13}$ & $20.1_{-0.2}^{+0.3}$ \\
LowDen & $0.667_{-0.133}^{+0.133}$ & $0$ & $1.17_{-0.23}^{+0.17}$ &$4.47_{-0.37}^{+0.37}$ \\
HighDen & $4.74_{-0.46}^{+0.46}$ & $0.0641_{-0.0641}^{0.0641}$ & $7.37_{-0.64}^{+0.71}$ &$52.4_{-1.2}^{+1.0}$\\
LowBin & $0.667_{-0.133}^{+0.133}$ & 0 & $0.967_{-0.167}^{+0.200}$& $8.87_{-0.50}^{+0.43}$\\
HighBin & $3.77_{-0.37}^{+0.34}$ & $0.100_{-0.067}^{+0.039}$ & $6.23_{-0.37}^{+0.50}$&$29.6_{-0.8}^{+0.7}$ \\
\hline
\end{tabular}
\end{table*}

Table \ref{tab-res} shows that about 24\% of the planets are ejected for both the Jupiter and the JupEnT runs. This can be compared to the simple predictions from Equation \eqref{eq-scat-rate}. That equation, when integrated over time, prescribes the chance that an event happens for a planetary system, if knowing the respective cross section $\sigma$. A number of authors have estimated that for EJEC under different assumptions \citep[e.g.,][]{Laughlin1998,Adams2006,Li2015,Wang2020}; here we take the value from \citet{Li2020c} where the setup was the most similar to this work. From there, $\sigma_\mathrm{EJEC}=9.7\times10^4$ au$^2$ implies a percentage of 12\% for EJEC in 1 Gyr for Jupiter's ejection. So the two differ by a factor of two. \citet{Li2020c} also measured the $\sigma$ for the Sun-Jupiter pair to acquire a companion star and the inference was that almost all are expected to have a companion within 1 Gyr. Here we find that 46\% of the planetary systems in the Jupiter run obtain at least companion at some point in the simulation. Therefore, the percentages in this work agree with the expectations reasonably well.

Considering the HJs in the Jupiter run, the percentage of HJ\_ZKL is 2.4\% and that of HJ\_SCAT is a hundred times smaller. So the formation of HJ as a direct result of a scattering is extremely rare \citep{Hamers2017a}. Compared to HJ\_ZKL, a significantly larger proportion, 5.8\% end up as COL\_ZKL, meaning that in many cases, the ZKL cycle is not quenched by GR or tides. The temporal evolution of the percentages will be deferred to Section \ref{sec-dep} where we derive their time dependence.

In our treatment of tides, Equation \eqref{eq-tid-q} prescribes how $Q_\mathrm{tide}$ varies depending on the planet orbit \citep{Beauge2012} and is a fit to the model of \citet{Ivanov2011}. Taking  our nominal simulation as an example, the minimum possible planetary $Q_\mathrm{tide}$, corresponding to the most efficient tidal damping, is achieved when the planet is just touching the surface of the central host and is about 2000 for $a_\mathrm{pl}=5$ au. However, works looking into different modes have suggested that for extremely eccentric orbits, the equivalent $Q$ can be much smaller, possibly reaching a few tens or even a few \citep[e.g.,][]{Wu2018,Yu2022}. With a more efficient tidal model, planets of the fate COL\_ZKL may end up HJ\_ZKL.

Figure \ref{fig-aei_1_2} shows such an example. The initial conditions of the planetary system as well as the sequences of the stellar scatterings are exactly the same for the two panels. As we discussed earlier, when $Q_\mathrm{tide,pl,0}=5\times10^6$ (top panel), tidal dissipation is not fast enough and the ZKL effect goes unsuppressed and forces the planet onto the star (COL\_ZKL). When $Q_\mathrm{tide,pl,0}=5\times10^5$ (bottom panel), tides efficiently shrinks the planet's orbit, detaches it from the companion star, and hence stops further eccentricity excitation by the ZKL mechanism and an HJ forms (HJ\_ZKL).

As Table \ref{tab-res} shows, for the JupEnT run, the percentage for HJ\_ZKL and CKL\_ZKL are almost the same, both about 4.2~\%, so the creation of HJ\_ZKL is boosted by 70~\%. But the sum of HJ\_ZKL and CKL\_ZKL is 8.3~\% which is in excellent agreement with the Jupiter run, a phenomenon seen also in \citet{Petrovich2015,Anderson2016,Munoz2016}. In the JupEnT set, the percentage of HJ\_SCAT and EJEC are not affected by enhanced tides, both only related to the scattering process.

Additionally, about 1.5\% of the planets collide with their host star during the scattering and 1.2\% acquire orbits bound to the scatterer. We have omitted discussion on these two states as they will not affect the creation of HJs. But we note that both percentages are roughly a tenth of that of EJEC, consistent with the ratios of their respective cross sections as derived in \citet{Li2020c}.

\section{Population synthesis}\label{sec-pop}

In the previous section, we have shown with concrete examples that HJs may form via high-$e$ migration initiated by a companion star that the planetary host star acquires during a binary--single scattering in a stellar cluster. In this section, we perform sets of population synthesis simulations and explore the dependence of the efficiency of this mechanism on the properties of the cluster and the planetary system.

\subsection{Simulation parameters}

We fix the central host to be the Sun and the planet's physical parameters to be those of Jupiter. For all the runs, the tidal model has been the normal one \eqref{eq-tid-q} and no enhancement is effected. The planet's orbital distribution as we detail below is also the same for all following runs.

\begin{figure}
\includegraphics[width=\columnwidth]{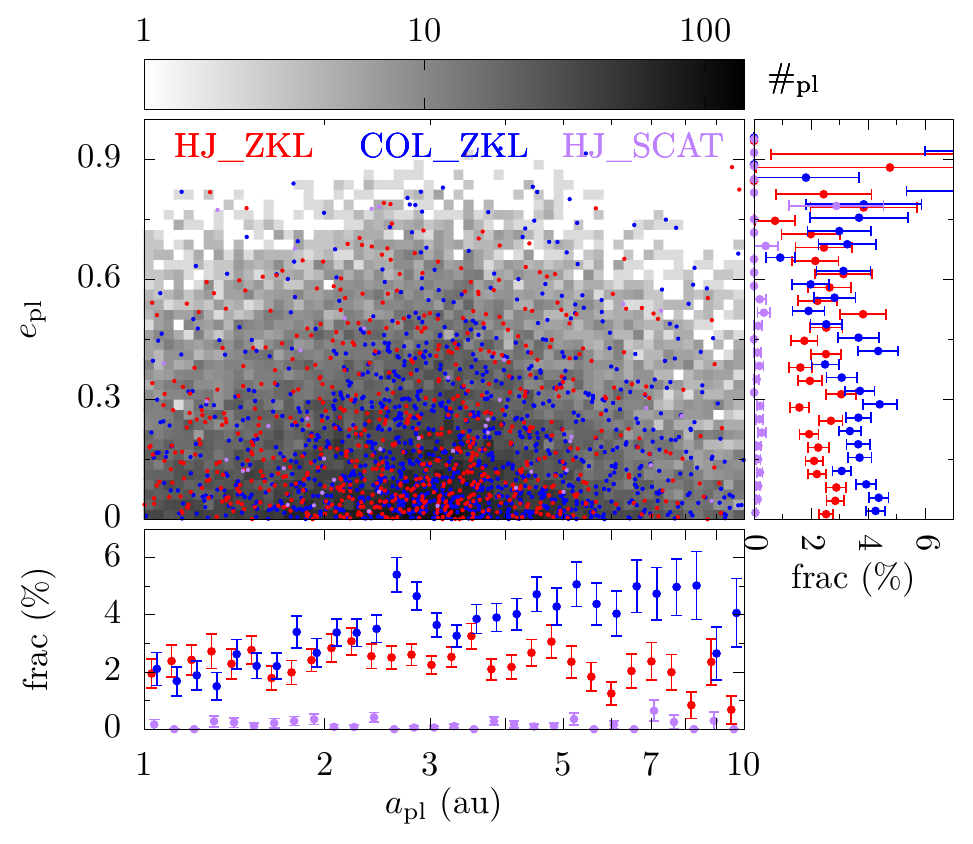}
\caption{The initial orbital distribution of the planets and their fates for the Nominal setup. The grey histogram in the big panel shows the planets' distribution in the $a_\mathrm{pl}-e_\mathrm{pl}$ plane, darker colours meaning more planets, as shown in the colour bar above. In that panel, the scattered points show those that have the fates HJ\_ZKL (red), COL\_ZKL (blue), and HJ\_SCAT (purple). The bottom and the right panels show the percentage of planets with those fates as a function of the initial orbit; the error bars are 1-$\sigma$ dispersion from a bootstrapping process; the points are slights shifted for better presentation.}
\label{fig-aeijup}
\end{figure}

We take the orbital parameters from the observed population. The distribution of the planet's orbital period $P$ follows a broken power law as derived in \citet{Fernandes2019} for radial velocity planets
\begin{equation}
\mathrm{PDF}(P) \propto \begin{cases} 
\left({P/P_\mathrm{b}}\right)^{p_{1}} & \mathrm{if} P \leq P_\mathrm{b}\\
\left({P/P_\mathrm{b}}\right)^{p_{2}} & \mathrm{if} P > P_\mathrm{b}.
\end{cases}
\label{eq-period}
\end{equation}
Here $P_\mathrm{b}=2075$ d, $p_1=0.7$ and $p_2=-1.2$. The inner boundary is 1 au as for closer-in planets, the ZKL timescales for the typical companion orbits from binary-single exchange are longer than those of GR/tides (see Figure \ref{fig-timescales}) so $e_\mathrm{pl}$ cannot be excited to values high enough to initiate efficient tidal damping. The outer boundary is somewhat arbitrary and we just let it be 10 au. The observed population of wide-orbit ($>$ 10 au) exoplanets is sparse and the errorbar in their distribution is large \citep[e.g.,][]{Nielsen2019,Wagner2022}. The grey histogram in Figure \ref{fig-aeijup} shows the initial orbital distribution of the planet.

Our eccentricity distribution follows a Beta distribution as proposed by \citep{Kipping2013a} for radial velocity planets. \footnote{Random number generators by Richard Chandler and Paul Northrop have been used \url{https://www.ucl.ac.uk/~ucakarc/work/randgen.html}.} An upper limit of $e_\mathrm{pl}=0.95$ is set, as this coincides roughly with the highest observed eccentricity among the radial velocity planets \citep[e.g., HD 20782 b, though in a very wide binary system; see][]{Jones2006} and also insures that initial tidal effect is negligible even for $a_\mathrm{pl}=1$ au.

In one set of the runs, the cluster parameters are the same as the Jupiter run, i.e., $n_*=50$ pc$^{-3}$ and $f_\mathrm{bin}=0.5$. This forms our main simulation set and is called the ``Nominal'' set. A total of 30000 runs are done for this set.

Four additional sets of simulations with different cluster properties are performed. In the sets ``LowDen''and ``HighDen'', $n_*=10$ and 200 pc$^{-3}$, respectively, both with $f_\mathrm{bin}=0.5$. And in the sets ``LowBin'' and ``HighBin'', $f_\mathrm{bin}=0.1$ and 0.9, respectively, both with $n_*=50$. For those, 3000 runs are done for each. These parameters are listed in Table \ref{tab-set}.

\subsection{Results of the Nominal simulation set}
We first analyse the results of the Nominal set. Table \ref{tab-res} shows that the percentage of HJ\_ZKL is 2.4\%, that of HJ\_SCAT 0.13\%, 3.6\% for COL\_ZKL, and 20\% for EJEC. In comparison to the Jupiter run, it seems that the change is mild -- the creation of HJ\_ZKL has the same efficiency and that of EJEC decreases by less than 20\%; HJ\_SCAT is enhanced by a factor of a few but its contribution to the formation of HJs is anyway less efficient by a factor of at least 20 compared to HJ\_ZKL. For COL\_ZKL, there is a 40\% boost (at $\sim5-\sigma$ level) in the Nominal run compared to the Jupiter run.

How does the planet's initial orbit affect its fate? The large panel of Figure \ref{fig-aeijup} displays as scattered points the initial orbital distribution of the planets in the final states HJ\_ZKL (red), COL\_ZKL (blue), and HJ\_SCAT (purple). The bottom and the right panels of that figure present the percentage of planets with the three fates as a function of $a_\mathrm{pl}$ and $e_\mathrm{pl}$.

The figure suggests that HJ\_ZKL does not depend on the initial $e_\mathrm{pl}$ and seems to show a weak negative dependence on $a_\mathrm{pl}$ \citep[e.g.,][]{Munoz2016}. The planet's $a_\mathrm{pl}$ affects the planet's evolution in many aspects. Figure \ref{fig-timescales} shows that for a fixed $a_\mathrm{com}$, a larger $a_\mathrm{pl}$ means a smaller ZKL timescale, facilitating HJ\_ZKL. But this could turn out to be an adverse effect as the ZKL effect could go untamed (by GR/tides) so that the planet collides with the central host. On the other hand, efficient tidal dissipation has to be activated so the planet's orbit can be shrunk. From our tidal model, this means $e_\mathrm{pl}>1-0.022\,\mathrm{au}/a_\mathrm{pl}$. Apparently, the larger the $a_\mathrm{pl}$ the higher the $e_\mathrm{pl}$ is needed; this works against HJ\_ZKL for a larger $a_\mathrm{pl}$. Moreover, embedded in a cluster, the constant stellar scatterings may alter the companion star's orbit and therefore interrupt the ZKL cycle. Overall, HJ\_ZKL shows a weak negative dependence on $a_\mathrm{pl}$ while for COL\_ZKL, a clearer positive dependence is seen.

Similarly, the effect of the initial $e_\mathrm{pl}$ on HJ\_ZKL is weak. This is seemingly counter-intuitive since a larger initial $e_\mathrm{pl}$ reduces the requirement on $i_\mathrm{pl,com}$ to excite $e_\mathrm{pl}$ to the same level \citep[e.g.,][]{Li2014a}. Take $a_\mathrm{pl}=5$ au for example, $e_\mathrm{pl}$ has to reach 0.996 to enable tidal dissipation (so $r_\mathrm{peri,pl}=0.022$ au). Using the leading order ZKL theory, we have performed a simple Monte Carlo simulation fixing the initial $e_\mathrm{pl}$ and randomly drawn $i_\mathrm{pl,com}$ and phase angles and the fraction of orbits that can achieve a maximum $e_\mathrm{pl}$ of at least 0.996 is calculated. We find that this fraction depends on $e_\mathrm{pl}$ very mildly and an increase of the initial $e_\mathrm{pl}$ from $\sim0$ to $\sim0.9$ only boosts the fraction by 100\%. But planets with initial $e_\mathrm{pl}\gtrsim0.9$ are rare in our simulations. This seems at odds with \citet{Mustill2022}. In explaining the observed high eccentricity of HR5183b, \citet{Mustill2022} found that an initial $e_\mathrm{pl}$ (which might be caused by planet--planet scattering) would enhance the chance that a companion excites $e_\mathrm{pl}$ to the observed value. In that work, the authors were examining the fraction of time that $e_\mathrm{pl}$ is higher than a certain value whereas here it is the maximum $e_\mathrm{pl}$ ever attained that matters.

\begin{figure}
\includegraphics[width=\columnwidth]{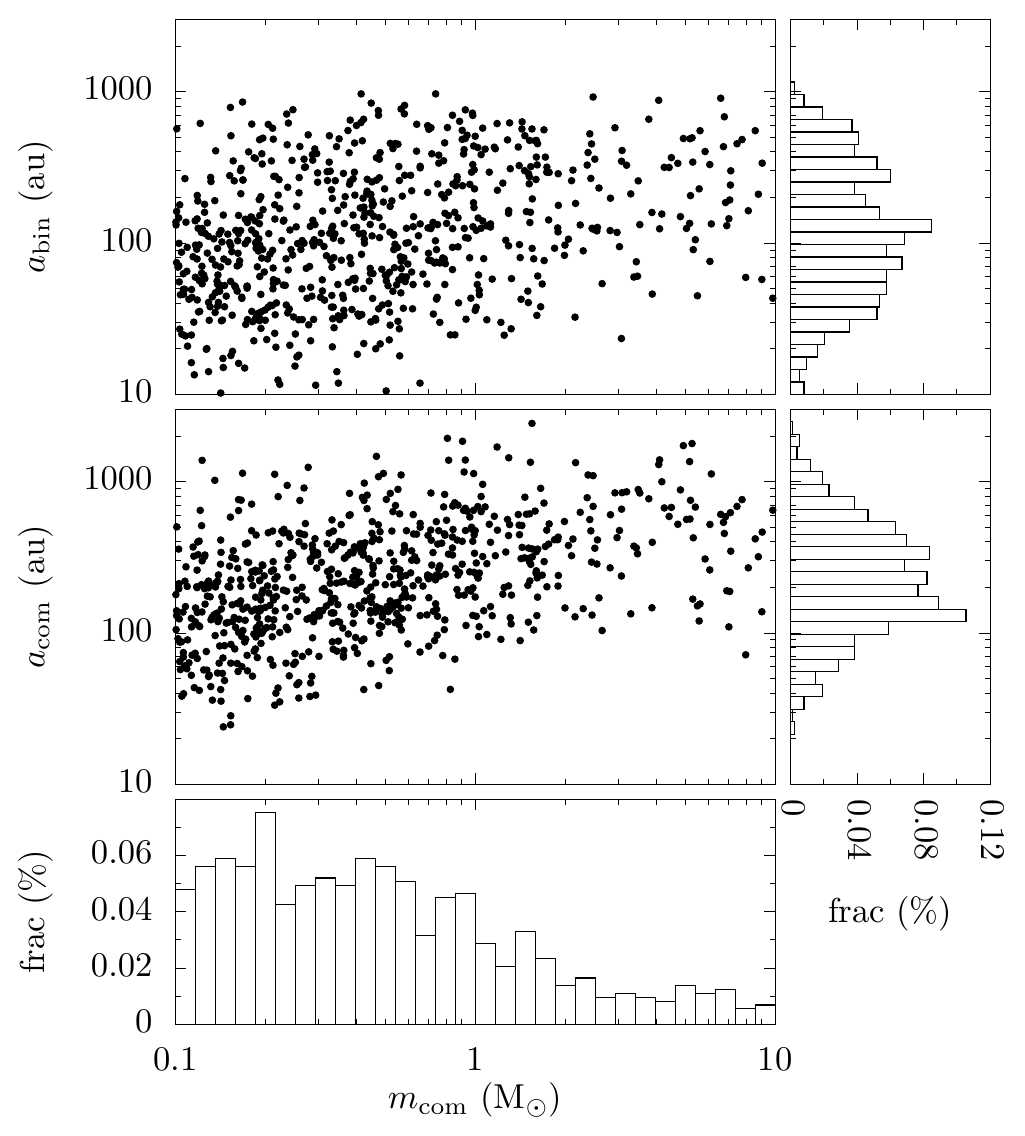}
\caption{The distribution of the scattering binary's $a_\mathrm{bin}$ leading to the outcome of HJ\_ZKL and the planetary system's companion's $a_\mathrm{com}$ as a function of the companion mass $m_\mathrm{com}$ in the nominal set. From top to bottom, the three histograms show the distribution of $a_\mathrm{bin}$, $a_\mathrm{com}$, and $m_\mathrm{com}$.}
\label{fig-hj_com}
\end{figure}

We would like to assess what kind of scattering binaries help create HJ\_ZKL and how the properties of the companions are distributed for these systems. But this may not be as straightforward as it may seem. As Figure \ref{fig-aei_105_1} shows, the orbit of the planetary system's companion is subject to further alteration due to stellar scattering. In that example, the companion star has remained bound to the planetary system so the binary scatter that this companion is in originally is the one that contribute directly to HJ\_ZKL. But it can be more complicated in that we have registered in our simulations cases where the planetary system obtains a companion star after scattering with a binary and an HJ does not form; after interactions with other scatterers (single or binary), the original companion swaps with another star and this companion triggers the process of HJ\_ZKL. In this latter case, if the companion at the time of the formation of the HJ comes from a binary scatterer, we record the orbit of that binary; if not, we trace back to see if the predecessor of the companion is from a binary and so on and so forth. The top left panel of Figure \ref{fig-hj_com} shows the semimajor axis of the scattering binary $a_\mathrm{bin}$ as a function of the mass $m_\mathrm{com}$ of the companion of the planetary system. The top right and the bottom panels show the histogram of $a_\mathrm{bin}$ and $m_\mathrm{com}$, respectively. The mass $m_\mathrm{com}$ covers the full range of our initial mass function with a median of 0.42 M$_\odot$ while $a_\mathrm{bin}$ is broadly distributed from tens to hundreds of au with a median of 110 au. The middle left panel of the figure shows the companion's $a_\mathrm{com}$ as a function of $m_\mathrm{com}$ and the middle right the histogram of $a_\mathrm{com}$. Compared to the broad distribution $a_\mathrm{bin}$, $a_\mathrm{com}$ is more centred around the median 220 au.

Many workers have carried out population synthesis studies on HJ formation via the ZKL mechanism \citep{Naoz2012,Petrovich2015,Anderson2016,Munoz2016,Vick2019}. Due to the inherently different assumptions (like the usage of secular/full equations of motion, the tidal model, and the cluster environment), our HJ formation rate HJ\_ZKL cannot be compared directly with theirs. But many common characteristics are observed:  e.g., the general HJ\_ZKL rate of a few per cent and the invariability of the sum of the rates of COL\_ZKL and HJ\_ZKL under different tidal efficiencies \citep[see Table \ref{tab-res} and ][]{Petrovich2015,Anderson2016,Munoz2016}. And the cluster environment also introduces new features. For instance, the preferred companion separation for HJ\_ZKL here under stellar scatterings (a few hundreds of au) is appreciably smaller than if the systems are in isolation \citep[wider than several hundreds of au][]{Naoz2012,Petrovich2015} as a result of the disruption of the companion orbit.

\subsection{Results of the other runs}
The percentages of different fates for the other simulation sets are presented in Table \ref{tab-res}.

First, what role does the stellar number density $n_*$ play? By comparing the sets LowDen and HighDen with the Nominal simulation, we observe that decreasing/increasing $n_*$ has the effect of counteracting/boosting the percentage of HJ\_ZKL. Obviously, a lower/higher $n_*$ implies a lower/higher scattering rate, which diminishes/enhances the chance that a planetary system acquires a companion star and therefore the probability of HJ\_ZKL. Also, a lower/higher $n_*$ implies a longer/shorter lifetime of the so-acquired companion star, allowing the ZKL effect more/less time to operate. According to Figure \ref{fig-timescales}, even for the HighDen run, for any $a_\mathrm{com}\lesssim$ 1000 au, the ZKL timescale is much shorter than the companion lifetime. Therefore, the constraint from the companion star's survivability is weak \citep[but higher order effects, not reflected in that figure, can operate on much longer timescales; see e.g.,][]{Ford2000,Naoz2011,Antognini2015} and as a consequence, for the parameter range considered here, a higher $n_*$ means a higher percentage of HJ\_ZKL. But the dependence may not be linear. Comparing the Nominal with the HighDen run, an increase of $n_*$ from 50 to 200 pc$^{-3}$ only increases the percentage of HJ\_ZKL by a factor of two, the main reason being that 1.5 times more planets are ejected in the denser environment so the reservoir for HJ\_ZKL is significantly smaller. A comparison between the Nominal and the LowDen runs shows that decreasing $n_*$ by 80\% leads to a drop in the percentage of HJ\_ZKL by more than 70\%, so the linearity towards smaller $n_*$ is more pronounced.

Much like the influence of $n_*$, $f_\mathrm{bin}$ affects a planetary system in two ways: increasing the prospect of the acquisition of a companion star and decreasing the lifetime of the companion. It turns out that a higher $f_\mathrm{bin}$ (HighBin) gives rise to a higher percentage of HJ\_ZKL and vice versa (LowBin). We note that in the LowBin run, the effective density for binaries ($n_\mathrm{bin}=n_*f_\mathrm{bin}$) is 50 pc$^{-3}\times0.1=5$ pc$^{-3}$ coincident with that in the LowDen run 10 pc$^{-3}\times0.5=5$ pc$^{-3}$, and the percentages of HJ\_ZKL are in excellent agreement in the two sets of simulations. This suggests that (when EJEC is not overwhelming) the percentage of HJ\_ZKL depends on the binary spatial density $n_\mathrm{bin}$ of the cluster only.

Then, not surprisingly, in all simulations, the percentage of HJ\_SCAT is smaller than that of HJ\_ZKL by at least an order of magnitude so we omit discussion on the former. And for all these simulation sets, the ratio of the percentage of COL\_ZKL and HJ\_ZKL is about constant $\sim1.5$, consistent with the expectation both are results of the ZKL mechanism and depend on the cluster property in similar ways. Therefore, we may broadly refer to both COL\_ZKL and HJ\_ZKL as ACT\_ZKL, meaning that extreme ZKL cycles are activated where the planet either turns into an HJ or plummet into the central host, i.e., ACT\_ZKL=COL\_ZKL+HJ\_ZKL.

\subsection{Empirical dependences on cluster parameters}\label{sec-dep}
In general, the rate that an event happens can be estimated with Equation \eqref{eq-scat-rate} by plugging in the appropriate $\sigma$. In calibrating $\sigma$, previous works have often separated the effects of binary and single stars \citep{Adams2006,Li2015,Li2020c}. Here we follow the same approach.

In our scenario, ACT\_ZKL is only affected by the binaries and single stars cannot contribute. So the rate of ACT\_ZKL can be approximated by $A_\mathrm{z} {n_\mathrm{bin}\over 1\, \mathrm{pc}^{-3}}$ where $A_\mathrm{Z}$ is a constant to be determined.

Apparently, ACT\_ZKL may only occur for planet that is still revolving around the host star (excluding those turning into HJs already). The size of this reservoir is declining because of ACT\_ZKL itself, ejection, and capture and collision during the scattering (the latter two are minor and are not discussed in detail in this work). Suppose the rate of all these effect combined is $A_\mathrm{r} {n_\mathrm{bin}\over 1\, \mathrm{pc}^{-3}}+B_\mathrm{r} {n_\mathrm{sin}\over 1\, \mathrm{pc}^{-3}}$ (where $A_\mathrm{r}$ and $B_\mathrm{r}$ are constants). Then the percentage of the size of the reservoir at time $t$ compared to the initial size is
\begin{equation}
e^{-(A_\mathrm{r} {n_\mathrm{bin}\over 1\, \mathrm{pc}^{-3}}+B_\mathrm{r} {n_\mathrm{sin}\over 1\, \mathrm{pc}^{-3}}){t\over 1\,\mathrm{Myr}}}\times100\%,
\end{equation}
where $t$ is the current time. Therefore, the rate of ACT\_ZKL at $t$ is 
\begin{equation}
{\mathrm{d}\mathrm{ACT\_ZKL}\over \mathrm{d}t} =e^{(A_\mathrm{r} {n_\mathrm{bin}\over 1\, \mathrm{pc}^{-3}}+B_\mathrm{r} {n_\mathrm{sin}\over 1\, \mathrm{pc}^{-3}}){t\over 1\,\mathrm{Myr}}}A_\mathrm{z} {n_\mathrm{bin}\over 1\, \mathrm{pc}^{-3}}.
\end{equation}
When integrating from time 0 to $t$, the percentage of ACT\_ZKL as a function of time is
\begin{equation}
\label{eq-dep1}
\begin{aligned}
\mathrm{ACT\_ZKL}=&{A_\mathrm{z} {n_\mathrm{bin}\over 1\, \mathrm{pc}^{-3}} \over A_\mathrm{r} {n_\mathrm{bin}\over 1\, \mathrm{pc}^{-3}}+B_\mathrm{r} {n_\mathrm{sin}\over 1\, \mathrm{pc}^{-3}}}\\ 
&\times(1-e^{-(A_\mathrm{r} {n_\mathrm{bin}\over 1\, \mathrm{pc}^{-3}}+B_\mathrm{r} {n_\mathrm{sin}\over 1\, \mathrm{pc}^{-3}}){t\over 1\,\mathrm{Myr}}})\times100\%.
\end{aligned}
\end{equation}

\begin{figure}
\includegraphics[width=\columnwidth]{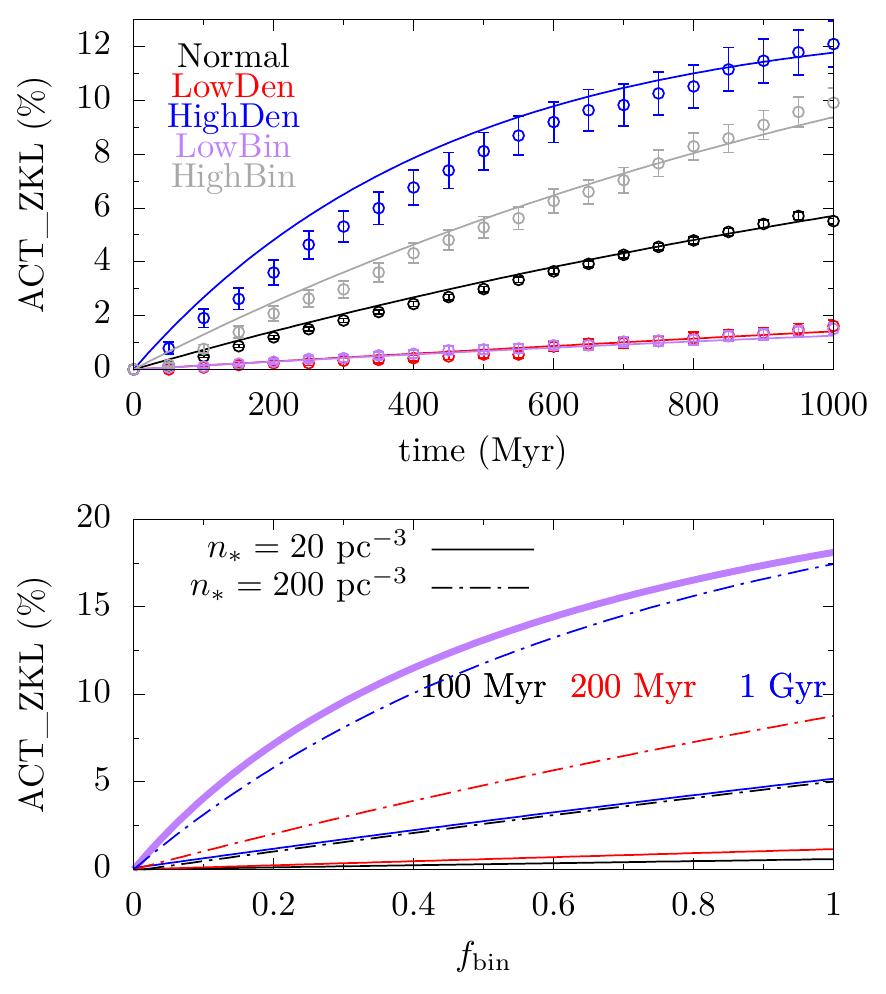}
\caption{Percentage of planets with the outcome ACT\_ZKL, the sum of HJ\_ZKL and COL\_ZKL. The top panel shows the time evolution of ACT\_ZKL from different runs (points) and the respective fits (line) in different colours. The bottom panel shows the percentage of ACT\_ZKL as a function of $f_\mathrm{bin}$ for $n_*=20$ (solid line) and 200 pc$^{-3}$ (dash-dotted line) at 100 Myr (black), 200 Myr (red) and 1 Gyr (blue). The thick purple line is the upper limit of the percentage for a given $f_\mathrm{bin}$.}
\label{fig-dependence}
\end{figure}

The top panel of Figure \ref{fig-dependence} shows the time evolution of the percentage of ACT\_ZKL for all the five population synthesis simulation sets. And we have fitted those curves using Equation \eqref{eq-dep1} above and the fitting parameters are $A_\mathrm{z}=(3.0\pm0.05)\times10^{-6}$, $A_\mathrm{r}=(1.6\pm0.2\times10^{-5}$, $B_\mathrm{r}=(6.3\pm1)\times10^{-6}$. The result from the fit is also presented. The agreement is fairly good and the largest deviation is within two sigma. It seems that while the percentage of ACT\_ZKL for the HighDen set is plateauing toward the end of the simulation, those for the other sets are still steadily increasing.

Now we rewrite Equation \eqref{eq-dep1} using $n_*$ and $f_\mathrm{bin}$ as
\begin{equation}
\label{eq-dep2}
\begin{aligned}
\mathrm{ACT\_ZKL}=&{A_\mathrm{z}f_\mathrm{bin} \over A_\mathrm{r}f_\mathrm{bin} +B_\mathrm{r}(1-f_\mathrm{bin}) } \\
&\times(1-e^{-[A_\mathrm{r}f_\mathrm{bin} +B_\mathrm{r}(1-f_\mathrm{bin})]{n_*\over 1\, \mathrm{pc}^{-3}}{t\over 1\,\mathrm{Myr}}})\times100\%.
\end{aligned}
\end{equation}
Equation \eqref{eq-dep2} shows that given enough time, the percentage of ACT\_ZKL will eventually get to an upper limit $A_\mathrm{z}f_\mathrm{bin}/[A_\mathrm{r}f_\mathrm{bin} +B_\mathrm{r}(1-f_\mathrm{bin})]\times100\%$ which is determined solely by $f_\mathrm{bin}$. As shown in the bottom panel of Figure \ref{fig-dependence}, this limiting value is increasing with $f_\mathrm{bin}$, reaching 10\% at $f_\mathrm{bin}=0.3$ and slowly levelling off toward 18\% at $f_\mathrm{bin}=1$.

The density $n_*$ prescribes how quickly the percentage of ACT\_ZKL approaches that limiting value. In the bottom panel of Figure \ref{fig-dependence}, we plot the percentage of ACT\_ZKL as a function of $f_\mathrm{bin}$ at 100 Myr (black), 200 Myr (red), and 1 Gyr (blue) for $n_*=$ 20 pc$^{-3}$ (solid line) and 200 $^{-3}$ (dash-dotted line). For the lower density, the percentages at all times are quasi-linearly dependent on $f_\mathrm{bin}$ but for the higher density, the percentage of ACT\_ZKL saturates toward the upper limit, implying that the limit can be reached within a few Gyr for $n_*$ of a few hundred pc$^{-3}$.

Finally, we note that a fraction of those ACT\_ZKL will be indeed HJ\_ZKL while the others will be COL\_ZKL. The exact division depends on the details of the tidal interaction; see the simulation sets Jupiter and JupEnT in Section \ref{sec-jup-res} for a discussion. But the chances for ACT\_ZKL and HJ\_ZKL are comparable.

\section{Discussion} \label{sec-dis}

\subsection{Observational implications}
As reviewed in the introduction section already, the observations of planets in clusters have been sparse with only a few HJs detected so far. Here we only discuss those found in dedicated surveys but not otherwise \citep[e.g.,][]{Obermeier2016,Ciardi2017,Rizzuto2018,Livingston2019}.

In total, 3 HJs have been found around 160 stars in Praesepe (NGC 2632) and Hyades \citep{Paulson2004,Quinn2012,Quinn2014} so the HJ occurrence rate is 2~\%. Both clusters are $\sim$ 600~Myr old and metal rich. After correcting for the solar metallicity, a rate of 1~\% was derived \citep{Quinn2014}, consistent with that of the field \citep{Wright2012}.

\citet{Brucalassi2016} surveyed 66 stars in M67 (NGC 6282) which is of solar metallicity and age. Three HJs were found so the ccurrence rate is 4.5\%; removing the 12 stars that are in binaries, the HJ ccurrence rate around single stars was 5.6\%. These numbers are much higher than that of the field \citep[e.g.,][]{Wright2012}. M67 has a high $f_\mathrm{bin}\sim 30-40\%$ on average but could be as high as 70~\% near the centre \citep{Davenport2010,Geller2021}. Being among the oldest open clusters, M67 is highly evolved. From $N$-body simulations producing predictions consistent with the observations, the cluster probably through its lifetime has lost the majority of its total mass and $n_*$ at the core has remained largely around 100 pc$^{-3}$ and $f_\mathrm{bin}$ has not evolved significantly either \citep{Hurley2005,Hurley2007}. Combined, this means that at the core (where solar mass stars sink to), $n_\mathrm{bin}$ is perhaps several tens to a hundred pc$^{-3}$, within the optimal range for the HJ\_ZKL production from the bottom panel of Figure \ref{fig-dependence}. If like the field, the primordial giant planet occurrence rate is 10-20~\% within a few au \citep[e.g.,][]{Cumming2008} at the core of M67, our mechanism would predict an HJ occurrence rate of 1-2~\%. But we note these inferences are to be treated with caution; see Section \ref{sec-cav} for a brief discussion.

Curiously, the sample of \citet{Brucalassi2016} contained 12 stars with companions but no planet was detected around those. This may seem to be at odds with our mechanism showing that when the HJ forms, there is likely a companion star. We remind that in their binaries, the change in radial velocity is at least 1.7 km~s$^{-1}$ within a few hundreds to a thousand days \citep{Pasquini2012}. Take a binary of a solar mass on a circular orbit of semimajor axis $a$ for example. The orbital velocity is $v\sim{30\,\mathrm{km\,s}^{-1}/\sqrt{a}}$ and angular velocity $\omega\sim{2\pi\,\mathrm{yr}^{-1}/\sqrt{a^3}}$. In an edge-on configuration, if the angle between the orbital velocity and line of sight is $\theta$, the radial velocity is $v_\mathrm{r}=v\cos\theta$ and its change after some time $T$, is $\delta v_\mathrm{r}=v\omega T\sin\theta>1.7\,\mathrm{km\,s}^{-1}$ according to \citet{Pasquini2012}. Substituting the respective values, the binaries in \citet{Brucalassi2016} have
\begin{equation}
a<11\sqrt{T\over1\,\mathrm{yr}}\sqrt{\sin\theta}\,\mathrm{au}\lesssim20\,\mathrm{au}.
\end{equation}
In our mechanism, HJs tend to form with companions of a few hundreds of au (Figure \ref{fig-hj_com}) not included by \citet{Brucalassi2016}; moreover, such companions may well be disrupted during the cluster evolution \citep[e.g.,][]{Parker2009}. Hence, it is no surprise that in the binary sample of \citet{Brucalassi2016}, no HJs were observed.


\subsection{Comparison with other mechanisms}
Several works have been dedicated to the formation of HJs in star clusters. \citet{Shara2016} used fully-fledged $N$-body cluster simulations to address this issue, propagating the evolution of massive $\sim2\times10^4$-member clusters with a binarity of 0.1 to a few Gyr. Their derived HJ formation rate is  0.4\% per star or 0.2\% per planet and suggested that maybe tripling the binarity could increase the formation rate by 200\%. \citet{Hamers2017a} examined how (multiple) stellar scattering helps create HJs via high-$e$ migration in globular clusters. Unlike in open clusters, in these densely-populated environment, ejection is likely to remove most of the planets \citep[e.g.,][]{Davies2005}. After a careful search, \citet{Hamers2017a} found that for an initial semimajor axis of a few au, the favourable stellar density for making HJs is a few times $10^4$ pc$^{-3}$. \citet{Wang2020a} investigated the long-term evolution of a two-planet system after stellar flybys, concluding that higher-$e$ migration could be triggered by interplanetary ZKL mechanism and/or pure scattering, the former more efficient when the two orbits are wide apart. \citet{Rodet2021a} looked at a similar two-object scenario but concentrated on the case of wide-separation orbits. More recently, \citet{Wang2022} found that a multi-planet system may gain enough angular momentum deficit such that the system may become unstable afterwards and one of the planets might become an HJ.

Due to the different assumptions, it is impossible to make a full comparison between these works and ours. We just make some comments below.

\citet{Hamers2017a} considered a single-planet system and omitted binary stars, so therein, the planet's small pericentre distance can only be achieved during the scattering, in some sense close to our case HJ\_SCAT. From our simulations, the rate is for HJ\_SCAT $\lesssim0.1\%$ for a typical open cluster setup. This implies for single-planet systems in open clusters, our HJ\_ZKL mechanism is the most efficient.

\citet{Shara2016,Wang2020a,Rodet2021a,Wang2022} have all considered multi-planet systems. In order for instability or ZKL effects to occur within the planetary system, significant orbital angular momentum must be extracted from the outermost object during the scattering and the closest distance between the scatter and the planetary system must be comparable to the size of the latter. Therefore, a ``close scattering'' for the planetary system is needed and the wider the planetary system, the more efficiently their models work. In contrast, in our model, the scattering occurs between a planetary and a stellar binary and the former, as a whole, exchanges with a component of the latter. Hence, the closest distance during the scattering only needs to be comparable to size of the stellar binary which is often much larger than that of the planetary system. In this sense, a ``close scattering'' for the planetary system is not needed and the system never experiences instantaneous orbital alterations during the scattering. Not relying on a wide planetary system, our mechanism probably works better for compact systems.

\subsection{Caveats}\label{sec-cav}

In this work, we have tracked the evolution of a one-planet system in an open cluster, simulating its scattering with single and binary stars using a simple Monte Carlo approach. In order to study the effect of our proposed formation mechanism for HJs, several potentially important factors have been omitted.

Open clusters, as the name suggests, are slowly losing their mass owning to member star ejection and stellar evolution \citep{Lamers2006}. In the meantime, cluster properties, like $f_\mathrm{bin}$ and $n_*$, may also evolve considerably over many-Myr timescales \citep[e.g.,][]{Kroupa1995}. Moreover, the parent cluster may be born with substructures where but these diffuse on many-Myr timescales \citep[e.g.,][]{Goodwin2004,Parker2012,Parker2012a}. Section \ref{sec-dep} suggests that the rate of ACT\_ZKL asymptotically approaches a value determined by the cluster's $f_\mathrm{bin}$ on a timescale typically of 1 Gyr. Therefore, the cluster's parameters used in this work are Gyr-averaged values.

The binary evolution has also been ignored. Wider binaries may be subject to disruption owing to stellar scattering \citep[e.g.,][]{Kroupa1995,Parker2009}. Figure \ref{fig-timescales} shows that those wider than a few hundreds of au would have been disrupted at a few hundreds of Myr so they cannot contribute to the formation of HJs at a later time. However, Figure \ref{fig-hj_com} shows that about half of the binary scatterers that lead to the formation of HJs via HJ\_ZKL have $a_\mathrm{bin}<100$ au and are largely immune from breakup. So the disruption of wide binaries in the cluster will potentially halve the HJ formation percentage we predict.

Then, the stellar evolution is also omitted. In a Gyr, a $\sim2$ M$_\odot$ star will evolve off the main sequence, shedding a large fraction of the initial mass \citep[e.g.,][]{Hurley2000}. If in a binary, this may cause the binary orbit to expand or even disrupt the binary totally \citep[e.g.,][]{Veras2011}. Figure \ref{fig-hj_com} shows that most of the companion stars (78\%) that contribute to HJ\_ZKL are below 1 M$_\odot$ and only 11\% of the companions are above 2 M$_\odot$. All such binaries are wide so when the massive companion is evolved, the Roche lobe will not be filled and the two stellar components evolve in isolation. Then as the stellar masses is being lost, the companion's orbit expands, making the ZKL timescale longer and itself vulnerable to scattering disruption and the outcome of HJ\_ZKL unlikely.\footnote{This is not like the case where if the planet-host is a massive star, its losing mass may enhance the ZKL effect \citep[e.g.,][]{Shappee2013,Stephan2021} and even lead to dynamical instability \citep[e.g.,][]{Kratter2012,Veras2017}.} Removing those stars, the percentage of HJ\_ZKL would drop by a few tens of per cent.
 
Studies of planets in clusters are limited by the relatively small number of stars in clusters compared to the field. Recently, \citet{Winter2020} calculated the phase space density for field stars using the full stellar kinematic information (position and velocity) and found the HJ occurrence rate was higher for stars in overdensities (which arise very much as a result of small relative velocities but not spacial proximity). Further analyses suggested that the multiplicity of a planetary system \citep{Longmore2021}, the architecture of multi-planet systems \citep{Chevance2021}, and the occurrence rates of some types of planets \citep{Dai2021} also have to do with the overdensities. If high phase space density now were to arise from a high-density birth environment, this would be a powerful tool to study the effects of birth environments on planetary system formation and early evolution. However, the statistical significance of these findings is questionable \citep{Adibekyan2021}. And it has been found that the stellar overdensities reflect the galactic kinematic/dynamical evolution \citep{Mustill2022a,Kruijssen2021} and are not necessarily relics of a clustered star formation. Therefore, we shy away from discussing the implications of these results on our result.

Finally, we have only examined a lone planet around a solar mass star. Statistically, how a multi-planet system evolves under stellar scattering depends on the architecture of the system, and instant instability, delayed instability may result \citep[e.g.,][]{Malmberg2011,Li2019,Li2020a,Wang2022}. If the system acquires a companion star, ZKL cycles/instability may be initiated or not also depending on the planets' configuration \citep[e.g.,][]{Innanen1997,Malmberg2007a,Marzari2022}. To present a thorough discussion on this is beyond the scope of this work.

\section{Conclusions} \label{sec-con}
We have proposed a formation channel for HJs in open star clusters: a planetary system, through binary-single interactions, acquires a companion star which then excite the planet's orbit through ZKL mechanism, activating high-eccentricity migration and giving rise to the creation of an HJ. Using Monte Carlo simulations, we have modelled how a solar mass star hosting a lone gas giant planet scatters with binary and single stars successively in an open cluster, tracking the evolution of the planet under Newtonian gravity, GR, and tides. Our main findings are as follows.
\begin{itemize}
\item If a solar mass star hosts a giant planet at a few au and acquires a companion star a few hundreds of au distant, that companion is able to excite the planet's orbit through ZKL mechanism, before it is stripped by stellar scattering in the cluster.
\item As a consequence, the planet's pericentre distance $r_\mathrm{peri,pl}$ may reach a few solar radii. If so, the planet's orbit can be shrunk by tidal dissipation in a few Myr and an HJ results.
\item In our nominal cluster with $n_*=50$ pc$^{-3}$ and $f_\mathrm{bin}=0.5$, $\sim2\%$ of single gas giants orbiting a solar mass star between 1 and 10 au will become an HJ through the above channel in a Gyr.
\item In the meantime, $\sim4\%$ of the planets collide with or are tidally disrupted by the host star because of the large-amplitude ZKL oscillations forced by the companion star.
\item And about 20\% of the planets are ejected from their host star owing to stellar scattering.
\item A far smaller percentage $\lesssim 0.1\%$ of the planets can acquire a small pericentre distance directly during stellar scattering and become HJs without the need of a companion star.
\item The total percentage of the formation of HJ and collision/tidal disruption depends on the cluster properties. The cluster $f_\mathrm{bin}$ sets an upper limit that will be reached given enough time ($10\%$ at $f_\mathrm{bin}=0.3$ and 18\% at $f_\mathrm{bin}=1$). And how quickly the above limit is reached depends linearly on $n_*$: a few Gyr for $n_*$ of a few hundred pc$^{-3}$.
\item Adopting a more efficient tidal model turns a fraction of the planets with the outcome collision into HJs. In general, the likelihoods of the formation of HJ and collision are comparable.
 
\end{itemize}

\section*{Acknowledgements}
The authors are grateful to the anonymous referee for the comments and suggestions that help improve the manuscript. The authors acknowledge financial support from the National Natural Science Foundation of China (grants 12103007 and 12073019) and the Swedish Research Council (grant 2017-04945) and the Swedish National Space Agency (grant 120/19C) and the Fundamental Research Funds for the Central Universities (grant 2021NTST08). This work has made use of the HPC facilities at Beijing Normal University.

\section*{Data Availability}

The data underlying this paper will be shared on reasonable request to the corresponding author.










\bsp	
\label{lastpage}
\end{document}